\documentclass[12pt]{article}
\usepackage{amsmath}
\usepackage{multirow}
\usepackage{amsthm}
\usepackage{amssymb}
\usepackage{amsfonts}
\usepackage{graphicx}
\usepackage{longtable}
\usepackage{enumerate}
\usepackage{enumitem}
\usepackage{natbib}
\usepackage{authblk}
\usepackage{threeparttable}
\usepackage{bm}
\usepackage{url} % not crucial - just used below for the URL
\usepackage{caption}
\newcommand{\blind}{1}

\newcommand{\R}{\mathbb{R}}
\newcommand{\norm}[1]{\left\lVert#1\right\rVert}

\newcommand{\E}{\mathbb{E}}

\newcommand{\setm}{\mathcal{M}}
\newcommand{\parenthese}[1]{\left(#1\right)}
\newcommand{\sparenthese}[1]{\Big(#1\Big)}
\newcommand{\bbrace}[1]{\left\{#1\right\}}

\DeclareMathOperator*{\argmax}{arg\,max}

\newcommand{\abs}[1]{\left\vert#1\right\vert}

\newcommand{\vx}{\bm{X}}
\newcommand{\vz}{\bm{Z}}
\newcommand{\vb}{\bm{\beta}}
\newcommand{\transpose}[1]{#1^T}
\usepackage{algorithm,algcompatible,amsmath}
\usepackage{algpseudocode}% http://ctan.org/pkg/algorithmicx
\algnewcommand\INPUT{\item[\textbf{Input:}]}%
\algnewcommand\OUTPUT{\item[\textbf{Output:}]}%

\newtheorem{theorem}{Theorem}

\newtheorem*{condition}{Condition}

\author[1]{Jie Zhou}

\newcommand\CoAuthorMark{\footnotemark[\arabic{footnote}]} % get the current value
\author[2]{Yue Zhang\protect\CoAuthorMark}
\author[2]{Zhangsheng Yu\thanks{ \texttt{yuzhangsheng@sjtu.edu.cn}; Corresponding author.}}
\affil[1]{%
	Department of Statistics, Shanghai Jiao Tong University, Shanghai
	200240, PR China}
\affil[2]{%
	Department of Bioinformatics and Biostatistics, Shanghai Jiao Tong University, Shanghai
	200240, PR China}

\begin{document}

\def\spacingset#1{\renewcommand{\baselinestretch}%
{#1}\small\normalsize} \spacingset{1}

%%%%%%%%%%%%%%%%%%%%%%%%%%%%%%%%%%%%%%%%%%%%%%%%%%%%%%%%%%%%%%%%%%%%%%%%%%%%%%

\if1\blind
{
%  \title{\bf Deep Partial Linear Cox Model}
    \title{\bf Partial Linear Cox Model with Deep ReLU Networks for Interval-Censored Failure Time Data}
%\author{
%%	Jie Zhou\hspace{.2cm}\\
%%	Department of Statistics, Shanghai Jiao Tong University\\
%%	Zhangsheng Yu\thanks{
%%		corresponding author} \\
%%	Department of Bioinformatics and Biostatistics, Shanghai Jiao Tong University}
%	Author 1\\}
\date{}
\maketitle
} \fi

%\if0\blind
%{
%  \bigskip
%  \bigskip
%  \bigskip
%  \begin{center}
%    {\LARGE\bf High Dimensional Functional Partially Linear Cox Model}
%\end{center}
%  \medskip
%} \fi

\bigskip
\begin{abstract} 
The partial linear Cox model for interval-censoring is well-studied under the additive assumption but is still under-investigated without this assumption. In this paper, we propose to use a deep ReLU neural network to estimate the nonparametric components of a partial linear Cox model for interval-censored data. This model not only retains the nice interpretability of the parametric component but also improves the predictive power compared to the partial linear additive Cox model. We derive the convergence rate of the proposed estimator and show that it can break the curse of dimensionality under some certain smoothness assumptions. Based on such rate, the  asymptotic normality and the semiparametric efficiency are also established. Intensive simulation studies are carried out to demonstrate the finite sample performance on both estimation and prediction. The proposed estimation procedure is illustrated on a real dataset.
\end{abstract}

\noindent%
{\it Keywords:}  deep neural network; partial linear Cox model; semiparametric inference; interval-censoring
\vfill

\newpage
\spacingset{1.5} % DON'T change the spacing!
\section{Introduction}
\label{sec:intro}

% introduction的写作思路
% 1. 区间删失是什么，重要性，例子
% 2. 

% 1. partial linear Cox model优势，介绍，成功应用
% 2. However, suffer from 维度灾难、
% 3. Deep network
% 4. We
% 5. 

In survival analysis, reliability studies and epidemiological studies, the accurate failure time is often not available but rather it is observed as an interval due to the regular follow-up. For example, in the Chinese Longitudinal Healthy Longevity Survey (CLHLS, \cite{Yi2008}), a subject's cognitive ability is measured by the Mini-Mental State Examination(MMSE, \cite{Katzman1988, Gagnon1990}) at each visit, and the cognitive impairment(CI) is defined by scores lower than a pre-specified threshold. Therefore, CI appeared in the interval formed by two consecutive visits. In statistics, this type of data is called interval-censoring. Another important application of interval-censored data is in the studies of transfusion-related acquired immune deficiency syndrome (AIDS), where the positive status of the human immunodeficiency virus (HIV) can never be known exactly, while the change in status is only known between some monitoring times.

Regression analysis of interval-censored failure time data aims to estimate the covariate effects on it. Semiparametric regression models with linear assumptions on the covariates have been widely developed including the Cox model(\cite{Finkelstein1986},\cite{Huang1996}), proportional odds model(\cite{HR}), accelerated failure time model(\cite{Rabinowitz1995, Betensky2001}), additive hazard model(\cite{Zeng, Wang2010}) and transformation model(\cite{Zhang2005, Wang2018}).

Although the linear assumption of covariate effects in the aforementioned models provides simplicity and interpretability,  this assumption is often violated in applications. For example, \cite{Lv2016} depicted the U-shaped relationship between blood pressure and the risk of CI in the CLHLS dataset. Partial linear models have been developed to model non-linear and linear effects simultaneously, e.g., \cite{Ma2005} first investigated a partial linear transformation model for current status data.  \cite{Cheng2011} extended the model to additively multivariate cases. \cite{Lu2018} and \cite{Liu2021} relaxed the the assumption of a stepwise constant baseline hazard function in these two models.

%Despite the progress achieved, further research on developing non-linear regression models to handle general dependence relationships between predictors and time events is needed.
%single covariate or additive

%The partial linear Cox model(PLC) is an extension of the linear Cox model(\cite{Cox1972}) and it is particularly well suited to study the highly nonlinear covariate effects on the failure time. For example, \cite{He2014} and \cite{Jørgensen2016} applied PLC to find that low and high  body mass index (BMI) is associated with elevated all-cause mortality. \cite{Du2009} also depicted the U-shaped relationship between white blood cells(WBC) count and the risk of diabetes with PLC.  A large literature has been established concerning the semiparametric inference of this model. \cite{Sasieni1992NonorthogonalPA} derived the efficient score function as well as the information bound.  \cite{Heller2001} proposed an inference procedure based on an orthogonal reparameterization of the partial likelihood. \cite{Cai2007} extended this model to multivariate survival outcomes. \cite{Jiang2011} provided a consistent variance estimator based on a bootstrap approach. \cite{Du2010} and \cite{Ma2012} considered the variable selection problem with fixed and diverging dimensions, respectively. \cite{Lu2018} put forward a flexible spline estimation method for PLC under monotonicity constraint.

%\cite{Therneau2000}
Nevertheless, none of the existing work has relaxed the additive assumption and allowed multivariate nonparametric function estimation in the partial linear Cox model for interval-censoring due to the curse of dimensionality. Intuitively, the overall convergence rate of these models decreases exponentially with respect to the dimension of the nonparametric component and will not be fast enough to establish the asymptotic normality for the parametric component(\cite{Shen1997}). 
% To alleviates curse of dimension one has to make assumption about the function space . 

Recently, deep neural networks(DNN, referred to as \emph{deep learning} in \cite{LeCun2015}) have emerged as a promising tool to alleviate the curse of dimensionality and have demonstrated superior performance for high-dimensional problems such as  image classification(\cite{NIPS2012_c399862d}), natural language processing(\cite{Devlin2019BERTPO}), and speech recognition(\cite{6296526}). Theoretical analysis attributes its success to its powerful ability to approximate functions from specific spaces such as the nested function space(\cite{Schmidt-Hieber2017, Bauer2019}) or the mixed smoothness Besov space(\cite{Suzuki2019})). For this reason, it has been extensively used in various types of survival models, such as  the Cox model(\cite{Katzman2018,Zhong2022}), the accelerated failure time model(\cite{Chen2019}), illness-death model(\cite{Cottin2022}), competing risk model(\cite{Lee2018}), cure rate model(\cite{Xie2021m,Xie2021p}), and nonparametric Cox model for interval-censoring(\cite{2206.06885, Sun2022}).

%However, research on semiparametric inference with DNN is extremely limited. \cite{Chen1999} derived the asymptotic normality and the efficiency bound for single-layer neural networks with possibly non-sigmoid activation functions, but is no longer valid for DNN and the rectified linear unit(ReLU) activation function(\cite{ReLU}). \cite{Crane-Droesch2017} and \cite{Zhong2021b} considered the semiparametric fixed effect model for panel data and the partial linear quantile regression model with deep ReLU network, respectively. For survival analysis, \cite{Xie2021m} and \cite{Xie2021p} studied the promotion and mixture cure rate model where the cure and survival component is estimated nonparametrically and parametrically each. None of these work examines the semiparametric efficiency theory with DNN.

In this paper, we propose the deep partial linear Cox model(DPLC)  for interval-censoring where covariates requiring  interpretability are kept linear and all the remaining covariates are modelled nonparametrically by a deep ReLU network. The proposed procedure enjoys some attractive properties. First, the convergence rate is justified to be free of the nonparametric dimension under some smoothness assumptions, suggesting that  the model can break curse of dimensionality.  Second, the parametric estimator  is shown to be asymptotically normal and  semiparametric efficient based on such rate. Third, a covariance estimator based on a least square regression rather than the computationally intensive bootstrap is provided to facilitate statistical inference.

 The rest of the paper is organised as follows. In Section~\ref{sec:meth}, we introduce notation, models, assumptions and the full likelihood of the observations for the interval-censored data. Section~\ref{sec:est} presents the estimation procedure based on the deep ReLU network.  The asymptotic properties of the proposed estimators are also discussed. In Section~\ref{sec:simu}, the finite-sample performance as well as the comparisons with other models are evaluated by a simulation study.  In Section~\ref{sec:app}, we apply the proposed model to two real datasets. Section~\ref{sec:conc} concludes with remarks and discussion.  Technical details of all the proofs are given in Supplementary Material.

\section{Model and Likelihood}
\label{sec:meth}

Suppose $\vx$ and $\vz$ are $\R^d$ and $\R^r$-valued covariates affecting the failure time $T$. We assume that given $\vx$ and $\vz$, $T$ follows the partial linear Cox model, e.g., its conditional cumulative hazard function has the form
\begin{equation}
\label{eq:model}
	\Lambda(t|\vx,\vz)=\Lambda_0(t)\exp\bbrace{\transpose{\vx}\vb+g(\vz)},
\end{equation}
where $\Lambda_0(t)$ is the unspecified baseline hazard function and $\bm{\beta}$ and $g$ correspond to the parametric coefficients and the nonparametric function, respectively. 

For interval-censored data, $T$ is known to lie within a random interval $(U, V)$, where $U$ and $V$ are the two examination times that satisfy $U<T\leq V$ with probability 1. We assume that $T$ is independent of $(U, V)$ given $(\bm{X}, \bm{Z})$ and the joint distribution of $(U, V)$ given $(\bm{X}, \bm{Z})$ is free of parameters $\theta=(\bm{\beta},\bm{\gamma})$. Let $\Delta_1 := 1_{\bbrace{T\leq U}}$, $\Delta_2 := 1_{\bbrace{U<T\leq V}}$, $\Delta_3:= 1_{\bbrace{T>V}}$, the observed information for a single object in the interval-censoring denoted by $O:=(\bm{X}, \bm{Z}, U, V,\Delta_1,\Delta_2,\Delta_3)$ is distributed as 
\begin{align*}
p(O)= (1-S(U|\vx,\vz))^{\Delta_1}\bbrace{S(U|\vx,\vz)-S(V|\vx,\vz)}^{\Delta_2}S(V|\vx,\vz)^{\Delta_3}p_{\vx,\vz}(U,V)h(\vx,\vz),
\end{align*}
where  $S(\cdot|\vx,\vz)=\exp(-\Lambda(\cdot|\vx,\vz))$ is the conditional survival function and $p_{\vx,\vz}(\cdot, \cdot)$ and $h(\vx,\vz)$ are the density functions of $(U, V)$ and $(\vx, \vz)$, respectively. Under the assumption that the distribution of $(U, V)$ is non-informative for $T$, then the log likelihood function of $\bbrace{(\bm{X}_i, \bm{Z}_i, U_i, V_i,\Delta_{1i},\Delta_{2i}\Delta_{3i}):i=1,..., n}$ is $l_n(\bm{\beta},\Lambda,g;\cdot):=\sum_{i=1}^n l(\bm{\beta},\Lambda, g; O_i)$ where
\begin{align*}
l(\bm{\beta},\Lambda, g; O)= \Delta_1\log(1- S(U|\bm{X}, \bm{Z}))+\Delta_2\log(S(U|\bm{X}, \bm{Z})-S(V|\bm{X}, \bm{Z}))+\Delta_3\log(S(V|\bm{X}, \bm{Z})).
\end{align*}

\section{Estimation and asymptotic properties}
\label{sec:est}

In this section, we consider the estimation of unknown parameters $(\bm{\beta},\Lambda_0, g)$.
A natural approach is to consider the sieve method(\cite{Chen2007}), e.g., maximizing $l_n(\bm{\beta},\Lambda_0, g)$ over the product space of $\R^d$ , sieve function spaces of $\Lambda_0$ and $g$ that grow in capacity with respect to $n$. We choose these two spaces as a monotone B-Spline space and a deep ReLU network space. 

The monotone B-Spline space $\setm$ is a non-negative linear span of the integrated spline  basis functions $M_k$(\cite{Ramsay1988}), i.e. $\setm = \{\gamma_kM_k:\gamma_k\ge 0\}$. Since each $M_k$
is non-decreasing function ranging from 0 to 1, constraining the coefficients to be non-negative guarantees the non-negativity and monotonicity of the estimator of the baseline cumulative hazard function $\Lambda_0$. As with B-splines, the $q_n = p_n + l$ basis functions are fully determined once the degree and the interior knot set are specified, where $p_n$ is the cardinality of the interior knot set.

A deep ReLU network space with input dimension $p_0$, depth $K$, hidden unit vector $\bm{p}=(p_0,\cdots,p_K,p_{K+1})$, sparsity constraint $s$ and norm constraint $D$ is defined as
\begin{align*}
\mathcal{G}(K, \bm{p}, s, D)=&\Bigg\{g(z)=(W^{(K)}\sigma(\cdot)+b^{(K)})\circ\cdots\circ (W^{(1)}z+b^{(1)}):\R^{p_0}\mapsto \R^{p_{K+1}},\Big.\\
&\quad\Big.\norm{g}_\infty\leq D, W^{(\ell)}\in\R^{p_{\ell+1}\times p_{\ell}}, b^{(\ell)}\in\R^{p_{\ell}+1},\ell=0,\cdots, K,\Big.\\
&\quad\Big. \sum_{\ell=0}^{K}\parenthese{\norm{W^{(\ell)}}_0 + \norm{b^{(\ell)}}_0}\leq s,\max_{\ell=0,\cdots, K}\norm{W^{(\ell)}}_\infty\vee\norm{b^{(\ell)}}_\infty\leq 1\Bigg\},
\end{align*} 
where $\norm{\cdot}_0$ and $\norm{\cdot}_\infty$ are the
$\ell_0$-norm and $\ell_\infty$-norm of a matrix, respectively and $W^{(\ell)}$ and $b^{(\ell)}$ are the weights and biases of the network, respectively. 

The estimation of $(\vb,\Lambda, g)$ is set to be the maximizer of $l_n$ over $\R^d\times \mathcal{M}\times \mathcal{G}(K, \bm{p}, s, \infty)$ with $p_0=r$ and $p_{K+1}=1$ after empirical centralization, that is
%\begin{equation}
\label{eq:est}
\begin{align*}
(\hat{\vb}_n,\hat{\Lambda}_n, \hat{g}_n)&=\argmax_{\R^d\times \mathcal{M}\times \mathcal{G}(K, \bm{p}, s, \infty)}l_n(\bm{\beta},\Lambda, g;\cdot)
%\hat{g}_n&= \hat{g}_n^{(u)}- 1/n\sum_{i=1}^n\hat{g}_n^{(u)}(\vz_i),\\
%\hat{\Lambda}_n&= \hat{\Lambda}_n^{(u)}\exp\sparenthese{1/n\sum_{i=1}^n\hat{g}_n^{(u)}(\vz_i)}.
\end{align*}
%\end{equation}

%\begin{align*}
%\hat{g}_n&= \hat{g}_n^{(u)}- 1/n\sum_{i=1}^n\hat{g}_n^{(u)}(\vz_i)\\
%\hat{\Lambda}_n&= \hat{\Lambda}_n^{(u)}\exp\sparenthese{1/n\sum_{i=1}^n\hat{g}_n^{(u)}(\vz_i)}
%\end{align*}

Such optimization poses several challenges. First, stochastic gradient descent(SGD, \cite{10.1214/aoms/1177729586}), the most widely used  algorithm in deep learning and its variants such as Adam(\cite{adam}) can not operate with the non-negativity constraint which is required on the coefficients of $\setm$. We remove the non-negativity constraint by reparametrization, that is, $\setm=\{\exp(\tilde{\gamma}_k)M_k:\tilde{\gamma}_k\in\R\}$. %the partial likelihood function is non-separable, i.e., it can not be formulated as an average of function values evaluated on the samples. Convergence properties of stochastic gradient descent(SGD, \cite{10.1214/aoms/1177729586}), the mostly used optimization algorithm in deep learning  for non-separable target functions are still unknown. We resort to the Adam(\cite{adam}) algorithm, a variant of SGD with batch size equals to $n$ to avoid this problem, i.e., no sampling strategy is used in this optimization.
The second challenge is regarding the specification of the hyperparameters of the network, mainly the depth $K$ and the knots $\bm{p}$. Unfortunately, there are no suitable model selection criteria for deep neural networks, such as  Akaike information criterion(AIC) or Bayesian information criterion(BIC)(\cite{Claeskens2001}) for linear sieve models. Meanwhile, cross-validation(CV), another popular method in model selection, is not applicable either due to the computational complexity of deep neural networks. Therefore, we select these hyperparameters in a lightweight and data-adaptive manner. A part of the dataset is randomly hold out as a validation set to select the hyperparameters among the candidates over $K$ and $\bm{p}$, and then the selected hyperparameters are used to rerun the estimation on the whole dataset.  The third challenge is to deal with the non-concavity with respect to the weights and biases of the network. We solve this problem to some extent by repeating this optimization for many times, each time with different initial values. Although the global maximizer is not reachable, this increases the probability of finding a better estimator. This claim is verified by numeric experiments later in Section~\ref{sec:simu}.
 
% variance estimator
%\begin{align*}
%\E \sup_{\eta/2\leq\norm{g-g_n}\leq \eta}\abs{\Mp_n(g)-\Mp_n(g_n)}
%\end{align*}

We now describe the asymptotic properties of the estimator $(\hat{\vb}_n,\hat{\Lambda}_n, \hat{g}_n)$. Denote  $\delta_n=\max_{i=0,\cdots,K}n^{-\bm{\alpha}_i/(2\bm{\alpha}_i+\tilde{\bm{d}}_i)}\vee n^{-\alpha_\Lambda/(2\alpha_\Lambda+1)}$ where $\bm{\alpha},\tilde{\bm{d}}$ and $\alpha_\Lambda$ is defined in the follow conditions.
\begin{condition}[C1]
	$\bm{\beta}$ belongs to a compact subset of $\R^{d}$
\end{condition}
\begin{condition}[C2]
	%	\begin{enumerate}[(1)]
	(1) $\E((\bm{X}, \bm{Z})^{\otimes 2})$ is positive definite.
	(2) both $\bm{X}$ and $\bm{Z}$ are bounded with probability 1.
	%	\end{enumerate}
\end{condition}

\begin{condition}[C3]
	The support of $U$ denoted as $[U_m, U_M]$ and that of $V$ denoted as $[V_m, V_M]$ satisify that $0=U_m<U_M=V_m<V_M$. The densities of $U$ and $V$ are both bounded away from zero and infinity on their support.
\end{condition}

\begin{condition}[C4]
	There exists some $\eta\in (0, 1)$ such that $\transpose{\bm{u}}\mathrm{Var}(\bm{X}|U)\bm{u} \ge  \eta \transpose{\bm{u}}\E(\bm{X}\transpose{\bm{X} }|U)\bm{u}$ and $\transpose{\bm{u}}\mathrm{Var}(\bm{X}|V)\bm{u} \ge  \eta \transpose{\bm{u}}\E(\bm{X}\transpose{\bm{X} }|V)\bm{u}$ for all $\bm{u}\in\R^{d}$ or $\transpose{\bm{u}}\mathrm{Var}(\bm{Z}|U)\bm{u} \ge  \eta \transpose{\bm{u}}\E(\bm{Z}\transpose{\bm{Z} }|U)\bm{u}$ and $\transpose{\bm{u}}\mathrm{Var}(\bm{Z}|V)\bm{u} \ge  \eta \transpose{\bm{u}}\E(\bm{Z}\transpose{\bm{Z} }|V)\bm{u}$ for all $\bm{u}\in\R^{r}$.
\end{condition}
%\begin{condition}[C3]
%	(a) These exist a $\xi_1>0$ such that $P(U-V\ge\xi_1) = 1$ (b)all possible conbinations of $(\bm{X}, \bm{Z})$ and $(\bm{\beta},\bm{\gamma})$ that makes $\lim_{t\rightarrow \infty}S_{(\bm{\beta},\bm{\gamma})}(t|\bm{X}, \bm{Z})=0$, there exists a $\xi_2 >0$ such that $S_{(\bm{\beta},\bm{\gamma})}(U|\bm{X}, \bm{Z})-S_{(\bm{\beta},\bm{\gamma})}(V|\bm{X}, \bm{Z})\ge\xi_2$ if $\Delta_2=1$, $1-S_{(\bm{\beta},\bm{\gamma})}(U|\bm{X}, \bm{Z})\ge\xi_2$ if $\Delta_1=1$, $S_{(\bm{\beta},\bm{\gamma})}(V|\bm{X}, \bm{Z})\ge\xi_2$ if $\Delta_3=1$
%	(c) 	the support of $U$ is $[0, U_M]$, the support of V is $[V_m, V_M]$ where $0<V_m<U_M<V_M$, density of $U$ and $V$ are bounded away from zero and infinity on their support respectively or $\norm{\cdot}_{L_2(U)}\asymp \norm{\cdot}_{L_2(\mu)}$ for Lebesgue measure $\mu$ on the support of $U$.
%\end{condition}

\begin{condition}[C5]
	$\Lambda_0 \in \mathcal{H}_1^{\alpha_\Lambda}([L, R], M)$, the H\"older space defined by 
	\begin{align*}
	\mathcal{H}_r^{\alpha}(\mathcal{D}, M)=\bbrace{g:\mathcal{D}\mapsto \R: \sum_{\kappa:|\kappa|<\alpha}\norm{\partial^{\kappa} g}_{\infty}+\sum_{\kappa:|\kappa|=\alpha}\sup_{x,y\in \mathcal{D},x\ne y}\frac{|\partial^{\kappa}g(x)-\partial^{\kappa}g(y)|}{\norm{x-y}_{\infty}^{\alpha-\alpha}}\leq M},
	\end{align*}
	and is monotonically increasing from $0$ with $\alpha_\Lambda \ge 2$.
\end{condition}

\begin{condition}[C6]
	$g\in \mathcal{H}(q,\bm{\alpha},\bm{d},\tilde{\bm{d}}, M)$, the composite smoothness function space(\cite{Schmidt-Hieber2017}) defined by
		\begin{align*}
	\mathcal{H}(q,\bm{\alpha},\bm{d},\tilde{\bm{d}}, M):=\{&g=g_q\circ\cdots\circ g_0: g_i=(g_{i1},\cdots,g_{id_{i+1}}),\\
	&g_i: \R^{d_i}\mapsto \R^{d_{i+1}}, g_{ij}\in \mathcal{H}^{\alpha_i}_{\tilde{d}_i}([a_i,b_i]^{\tilde{d}_i}, M)\}
	\end{align*}
	and $\E g(\bm{Z})=0$.
\end{condition}
\begin{condition}[C7] The hypyer-parameters of the deep ReLU network satisfies that
	$K=O(\log n)$, $s=O(n\delta_n^2\log n)$, $n\delta_n^2\lesssim \min_{k=1,\cdots,K}\bm{p}_k\leq \max_{k=1,\cdots,K}\bm{p}_k\lesssim n$ and $p_n=O(n^{1/(2\alpha_\Lambda+1)})$.
\end{condition}
Conditions (C1)-(C4) are similar to those in \cite{Zhang2010}. Condition (C5) and Condition (C6) restrict the function space of $\Lambda$ and $g$, respectively. Condition (C7) describes the growing of hyperparameters of DNN with respect to $n$.

We have the following theorems as $n\rightarrow \infty$.
\begin{theorem}[\bf{Convergence rate}]\label{thm1}	
	Under Conditions (C1)–(C7), the estimator $(\hat{\vb}_n,\hat{\Lambda}_n, \hat{g}_n)$ is consistent for $(\bm{\beta}_0, \Lambda_0, g_0)$ and the convergence rate is $\delta_n\log^2n$, e.g.,
	\begin{align*}
	||{\hat{\bm{\beta}}_n-\bm{\beta}_0}||_2+||\hat{\Lambda}_n-\Lambda_0||_{L_2(U, V)}+\norm{\hat{g}_n-g_0}_{L_2(\vz)}=O_P(\delta_n\log^2 n).
	\end{align*}
\end{theorem}

%\begin{theorem}[\bf{Efficient Score Function}]\label{thm2}	
%%\end{align}
%Under Conditions (C1)–(C7), the efficient score function for $\bm{\beta}$ is ${l}^*_{\bm{\beta}}(\bm{\beta},\Lambda,g; O)=\dot{l}_{\bm{\beta}}(\bm{\beta},\Lambda,g; O)-\dot{l}_1(\bm{\beta},\Lambda,g; O)[\bm{h}_1] -
%\dot{l}_2(\bm{\beta},\Lambda,g; O)[\bm{h}_2]$.
%\end{theorem}

\begin{theorem}[\bf{Asymptotic Normality and Efficiency}]\label{thm3}
	%	Let $V$ be the close linear span of $\mathcal{A}-\theta_0$ where $\mathcal{A}:=\bbrace{\theta\in\Theta:d_\Lambda(\theta,\theta_0)=O_P(\delta_n)}$, $\ip{v_1, v_2}:=-P\ddot{m}_{\theta_0}[v_1,v_2]$ for $v_1,v_2\in V$, $\mathcal{A}_n:=\Theta_n\cap \mathcal{A}$, $\theta_n:=\argmin_{\theta\in \Theta_n}\norm{\theta-\theta_0}$, $V_n:=\mathcal{A}_n-\theta_{n}$, suppose $\psi:\Theta\mapsto \R$ with a linear pathwise direcntional derivative $\dot{\psi}$, according to Reisz Representer Theorem, there exists a $v_n^*$ in $V_n$ such that for every $v \in V_n$
	%	\begin{align*}
	%	\dot{\psi}_{\theta_0}[v]=\ip{v_n^*,v}.
	%	\end{align*}

	Suppose Conditions (C1)-(C7) hold, $I_{\bm{\beta}}$ is nonsingular and $n\delta_n^4\rightarrow 0$, then 
	%	\begin{align*}
	%	\sqrt{n}\frac{(\psi(\thetahat_n)-\psi(\theta_0))}{\norm{v_n^*}^2}=\Gn\dot{m}_{\theta_0}[v_n^*/\norm{v_n^*}]+o_P(1)
	%	\end{align*}
	%	if futhermore, (C7)(i) or (C7)(ii) holds, then 
	%	\begin{align*}
	%	\sqrt{n}\frac{(\psi(\thetahat_n)-\psi(\theta_0))}{\norm{v_n^*}^2}\rightsquigarrow N(0, 1).
	%	\end{align*}
	%	As a result,
	\begin{align*}
	\sqrt n (\hat{\bm{\beta}}_{n}-\bm{\beta}_{0})\rightsquigarrow N(0, I^{-1}_{\bm{\beta}}),
	\end{align*}
	where  $I_{\bm{\beta}}$ is the semiparametric information bound for $\bm{\beta}$.
%	 defined by
%	\begin{equation}
%	I_{\bm{\beta}}=\E(\dot{l}_{\bm{\beta}}(\bm{\beta},\Lambda,g; O)-\dot{l}_1(\bm{\beta},\Lambda,g; O)[\bm{h}^*_1] - \dot{l}_2(\bm{\beta},\Lambda,g; O)[\bm{h}^*_2])^{\otimes 2}.
%	\end{equation}
\end{theorem}

It is interesting to notice that the polynomial term $\max_{i=0,\cdots,K}n^{-\bm{\alpha}_i/(2\bm{\alpha}_i+\tilde{\bm{d}}_i)}\vee n^{-\alpha_\Lambda/(2\alpha_\Lambda+1)}$ is free of the nonparametric dimension $r$ which means curse of dimensionality is much  eased in this model. Furthermore, the estimator of $\bm{\beta}$ attains the asymptotic normality and the semiparametric efficiency bound even though the overall convergence rate is slower than $n^{-1/2}$. 

To perform statistical inference on $\bm{\beta}_0$ with finite sample size according to Theorem~\ref{thm3}, one has to estimate the information matrix $I_{\bm{\beta}}$. Consider a parametric smooth submodel with parameter $(\bm{\beta}, \Lambda_{(s)},g_{(s)})$, where $\Lambda_{(0)}=\Lambda$, $g_{(0)}=g$ and $\frac{\partial}{\partial s} \Lambda_{(s)}=h_1,\frac{\partial}{\partial s} g_{(s)}=h_2$, the
%\begin{align*}
%\frac{\partial}{\partial s} \Lambda_{(s)}=h_1,\frac{\partial}{\partial s} g_{(s)}=h_2
%\end{align*}
score operators are defined by
\begin{align*}
\dot{l}_{\bm{\beta}}(\bm{\beta},\Lambda,g; o)&=\frac{\partial}{\partial \bm{\beta}} l(\bm{\beta},\Lambda_0,g; o),\\
\dot{l}_1(\bm{\beta},\Lambda,g; o)[h_1]&=\frac{\partial}{\partial s} l(\bm{\beta},\Lambda_{(s)},g; o)\big|_{s=0},\\
\dot{l}_2(\bm{\beta},\Lambda,g; o)[h_2]&=\frac{\partial}{\partial s} l(\bm{\beta},\Lambda_0,g_{(s)}; o)\big|_{s=0}.
\end{align*}
The \emph{least favourable direction} $(\bm{h}_1^*, \bm{h}_2^*)$ is obtained by projection $\dot{l}_{\bm{\beta}}$ into the product space of $\dot{l}_1$ and $\dot{l}_2$, or equivalently, minimize 
%\begin{align}
\begin{equation}
\rho(\bm{h}_1, \bm{h}_2)=\E||\dot{l}_{\bm{\beta}}(\bm{\beta},\Lambda,g; O)-\dot{l}_1(\bm{\beta},\Lambda,g; O)[\bm{h}_1] - 
	\dot{l}_2(\bm{\beta},\Lambda,g; O)[\bm{h}_2]||^2.
\end{equation}
Then $I_{\bm{\beta}}$ can be calculated by 
\begin{equation}
I_{\bm{\beta}}=\E[\dot{l}^*_{\bm{\beta}}(\bm{\beta},\Lambda,g; O)^{\otimes 2}]=\E[(\dot{l}_{\bm{\beta}}(\bm{\beta},\Lambda,g; O)-\dot{l}_1(\bm{\beta},\Lambda,g; O)[\bm{h}^*_1] - \dot{l}_2(\bm{\beta},\Lambda,g; O)[\bm{h}^*_2])^{\otimes 2}],
\end{equation}
where $\dot{l}^*_{\bm{\beta}}(\bm{\beta},\Lambda,g; O)$ is the efficient score function. Because there are no closed form expressions for $(\bm{h}_1^*, \bm{h}_2^*)$, following \cite{Huang2008}, we obtain  $\hat{\bm{h}}_1^*$ and $\hat{\bm{h}}_2^*$, the estimation of $\bm{h}_1^*$ and $\bm{h}_2^*$, by minimizing the empirical version of projection (2) over the product space of another two deep ReLU network spaces with properly chosen hyperparameters. The estimator of $I_{\bm{\beta}}$ is then defined by pluging $\hat{\bm{h}}_1^*$ and $\hat{\bm{h}}_2^*$  into the empirical version of (3).

%\begin{algorithm}
%	\caption{DPLCOX Algorithm}
%	\begin{algorithmic}[1]
%		\INPUT observations $(T_i,\Delta_i, \vx_i,\vz_i): i=1,\cdots, n$, network layers candidates $Ls$, network width candidates $Ds$
%		\OUTPUT $\hat{\vb}_n,\hat{\phi}_n, \hat{I}(\vb)$
%		\State split $(T_i,\Delta_i, \vx_i,\vz_i): i=1,\cdots, n$ into training set and validation set in 8:2.
%		\For{ $L$ in $Ls$}
%		\For{$D$ in $Ds$}
%		\For{$i$ in 1:6}
%		\State \parbox[t]{\dimexpr\textwidth-\leftmargin-\labelsep-\labelwidth}{maximize $l_n(\vb,\phi)$ with SGD composed of training set over $\R^d\times \Phi(L, D, \infty,\infty)$ \\and
%			evaluated in validation set, return likelihood in training set and validation\\
%			 set as $l^t_{L, D, i}$ and $l^v_{L, D, i}$ respectively.
%		}
%		\EndFor
%		\State $index=\argmax_{i=1}^6 l^t_{L,D,i}$, $l^{v*}_{L, D}=l^v_{L, D, index}$
%		\EndFor
%		\EndFor
%		
%		\State $L^*, D^*=\argmin_{L\in Ls, D\in Ds} l^{v*}_{L, D}$
%		\For{$i$ in 1:6}
%		\State{maximize $l_n(\vb,\phi)$ with SGD composed of whole dataset over $\R^d\times \Phi(L^*, D^*, \infty,\infty)$}
%		\State{set $(\hat{\vb}_n,\hat{\phi}_n)$ to correspond to maximal $l_n(\hat{\vb}_n,\hat{\phi}_n)$}
%		\EndFor
%		\State{minimize empirical version of \eqref{eq:info} composed of training set with early stopping keeping track of loss in validation set, return $\hat{a}^*,\hat{h}^*$}
%		\State{set $\hat{I}(\vb)$ to be empirical version of \eqref{eq:info2}}
%		\State {\textbf{Return} $(\hat{\vb}_n, \hat{\phi}_n)$ and $\hat{I}({\vb})$}
%	\end{algorithmic}
%\end{algorithm}

\section{Simulation}
\label{sec:simu}
In this section, we demonstrate the numerical performance of DPLC and compare it in terms of both estimation and prediction with linear Cox regression and partial linear additive Cox regression for interval-censoring. These two models are implemented using the R packages \emph{icenReg} and are abbreviated as CPH and PLAC, respectively. DPLC is implemented with the Python package \emph{pycox}  based on PyTorch(\cite{NEURIPS2019_9015}).

We first generate covariates $\vx$ and $\vz$ as follows
%\begin{equation*}
\begin{align*}
%\bm{U}_k&\sim U[0, 1],k=1,\cdots, 10\\
\vx&\sim \mathrm{Binomal}(1, 1/2)\\
\vz&\sim \mathrm{Clayton}(8, 0.5, [-2, 2])
\end{align*}
%\end{equation*}
$T$ is generated with $\Lambda_0(t)=\mu t^\kappa, \kappa\in (0.5, 1, 2)$ and $\vb_0=1.2$. After $T$ is generated, $(U, V)$ is obtained by dividing $[0, 5]$ into $10$ equal-distance intervals with visit probability $p\in (0.4, 0.7)$. $g$ is chosen from the following candidates:
\begin{enumerate}[font={\bfseries},label={Case \arabic*},wide=0pt, leftmargin=*]
	\item \textbf{(linear)}: $g(\vz)=2.4\sum_{k=1}^{10}\left(\vz_k-\frac 12\right)$;
	\item \textbf{(additive)}: $g(\vz)=1.2\sum_{k=1}^{10}\cos\left(\frac{2\pi}{k}\vz_k\right)$;
		\item \textbf{(deep-1)}: $g(\vz)=4.0\abs{\sum_{k=1}^{10}\left(\vz_k-\frac 12\right)}$;
			\item \textbf{(deep-2)}: $g(\vz)=4.5\left(\max_{k=1,2,3}\vz_k-\min_{i=1,2,3}\vz_k\right)$;
\end{enumerate}
Case 1 and Case 2 correspond to CPH and PLAC, respectively while Case 3 and Case 4 are designed for DPLC. The factors $2.4$, $1.2$, $4.0$ and $4.5$ in each case were used to scale $\text{Var}(\phi(\vz))/\text{Var}(\transpose{\vx}\vb)$ within the range $4-6$, i.e. to control the ratio of signals from the nonparametric and parametric components. Each dataset is randomly splitted with a 80:20 ratio as a training and validation set to select the best hyperparameters from all combinations of $K\in \{2, 3, 4\}$ and $\bm{p}_k \in \{\lceil u/4r\rceil:u=1,\cdots,8\}$. We repeat this 5 times with different initial values for the optimization in (\ref{eq:est}), with the chosen parameters corresponding to the maximal full likelihood in the validation set. 

% For each setting, $200$ replcates are producted with sample sizes $n\in \{1000, 2000\}$. Throughout the simulation, the data is splitted into training data and valudation data in a 80:20 ratio, another $4000$ samples are generated as test data.

The bias for the pamametric estimator $\hat{\bm{\beta}}_n$ and its empirical standard error from 500 replications with $n\in \{500, 1000\}$ are summarized in Table \ref{tbl:bias}. As expected, the bias of DPLC is comparable to, if not slightly worse than CPH and PLAC in Case 1 and Case 2, respectively, since these two cases are specifically designed for them. However, in Case 3 and Case 4, CPH and PLAC are more seriously biased than DPLC and do not improve with increasing $n$, whereas DPLC does. For example, in Case 3 with $n=500, p=0.5$ and $\kappa=1$, the bias of $\bm{\beta}_1$ for CPH and PLAC are $-0.782$ and $-0.412$, respectively, while the bias for DPLC is $-0.052$, a much smaller one and it decreases to $-0.037$ when $n$ increases to $1000$, but the bias for CPH and PLAC increases to $-0.812$ and $-0.438$, respectively. This phenomenon can be explained by the fact that the highly complicated nonparametric function $g$ in Case 3 and Case 4 can be easily fitted by a deep ReLU network, whereas it can not be approximated well by any linear or additive function,  and this inapproximability is further comfirmed with increasing $n$.  As might be expceted, empirical standard error decreases with increasing $n$ for all models and all censoring rates. Table \ref{tbl:cp} presents the converage proportion of the 95\% confidence intervals, suggesting that the empirical coverage probabilities for DPLC were generally around 95\% and close to the nominal level in four cases while those for Cox and PLAC are far away from 95\% in Case 3 and Case 4 due to the significant bias.

The performance in estimating of $g$ measured on a test data formed of $4000$ independent samples with the relative mean of squared error(RMSE) defined as
\begin{align*}
RMSE(\hat{g}_n)=\frac{\sum_{i=1}^n(\hat{g}_n(\vz_i)-g_0(\vz_i))^2}{\sum_{i=1}^n(g_0(\vz_i)-\bar{g}_0)^2},
\end{align*}
is reported in Table \ref{tbl:mse}. The smaller the metric is, the more accurate an estimator is. Similar to the results for the parametric estimator, DPLC significantly outperforms both Cox and PLAC by a lot in Case 3 and Case 4. To name a few, $0.448$ for DPLC versus $0.971$ and $0.936$ for CPH and PLAC in Case 4 with $n=500,p=0.4$ and $\kappa=0.5$. In Case 1 and Case 2, DPLC performs only slightly worse than CPH and PLAC.
We evalute and compare the predictive power of CPH, PLAC and DPLC with the Intergrated Mean Square Error(IMSE) in Table \ref{tbl:imse1} defined by
\begin{align*}
\mathrm{IMSE}&=\frac 1N\sum_{i=1}^N\sparenthese{\int_0^{U_i}(1-\hat{S}_n(t|\vx_i,\vz_i))^2dt+\int_{V_i}^\tau\hat{S}_n^2(t|\vx_i,\vz_i)dt}.
%\mathrm{IMSE}_2&=\frac 1N\sum_{i=1}^N\frac 1{\tau}\int_0^\tau \sparenthese{\hat{S}_n(t|\vx_i,\vz_i, U_i, V_i)-\hat{S}_n(t|\vx_i,\vz_i)^2}dt,
\end{align*}
It is can be seen from this table that the IMSE of DPLC is much smaller than that of CPH and PLAC  in Case 3 and Case 4. For example, it is $0.118$ for DPLC while is $0.173$ and $0.169$ for CPH and PLAC, respectively. In Case 1 and Case 2, CPH and PLAC outperform DPLC by only about $0.005$ IMSE. Further analysis of the simulation study can be found in Supplementary Material.

\begin{table}[]
			\caption{Bias and empirical standard error(in parentheses) of parametric estimator for the linear Cox model(CPH), partial linear additive Cox model(PLAC) and deep partial linear Cox model(DPLC)}
	\resizebox{\textwidth}{!}{
		\begin{tabular}{ccccccccccccc}\hline
			&      &     &      & \multicolumn{3}{c}{$\kappa=0.5$} & \multicolumn{3}{c}{$\kappa=1$} & \multicolumn{3}{c}{$\kappa=2$} \\\hline
			setting & $\rho$ & $p$ & $n$ &           CPH &     PLAC &     DPLC &        CPH &     PLAC &     DPLC &        CPH &     PLAC &     DPLC \\\hline
			linear & 0.5 & 0.4 & 500 &         0.064 &    0.168 &   -0.035 &      0.048 &    0.127 &   -0.097 &      0.084 &    0.161 &    -0.05 \\
			&      &     &      &       (0.320) &  (0.394) &  (0.368) &    (0.320) &  (0.372) &  (0.294) &    (0.356) &  (0.401) &  (0.358) \\
			&      &     & 1000 &         0.026 &    0.064 &    0.021 &      0.093 &    0.128 &   -0.004 &      0.072 &    0.099 &    0.002 \\
			&      &     &      &       (0.273) &  (0.294) &  (0.283) &    (0.233) &  (0.244) &  (0.231) &    (0.223) &  (0.228) &  (0.222) \\
			&      & 0.7 & 500 &         0.128 &    0.229 &   -0.074 &      0.063 &    0.121 &   -0.093 &       0.11 &    0.168 &    0.024 \\
			&      &     &      &       (0.393) &  (0.458) &  (0.378) &    (0.332) &  (0.366) &  (0.301) &    (0.261) &  (0.289) &  (0.265) \\
			&      &     & 1000 &         0.024 &    0.055 &   -0.079 &      0.043 &    0.073 &   -0.026 &      0.052 &    0.076 &   -0.009 \\
			&      &     &      &       (0.236) &  (0.254) &  (0.271) &    (0.216) &  (0.225) &  (0.213) &    (0.204) &  (0.208) &  (0.198) \\
			additive & 0.5 & 0.4 & 500 &        -0.443 &    0.186 &    -0.22 &     -0.422 &    0.114 &   -0.258 &     -0.519 &    0.074 &   -0.238 \\
			&      &     &      &       (0.289) &  (0.369) &  (0.328) &    (0.287) &  (0.316) &  (0.300) &    (0.223) &  (0.314) &  (0.281) \\
			&      &     & 1000 &        -0.524 &    0.037 &   -0.206 &     -0.508 &    0.029 &   -0.193 &     -0.481 &    0.105 &   -0.145 \\
			&      &     &      &       (0.185) &  (0.239) &  (0.222) &    (0.215) &  (0.246) &  (0.223) &    (0.186) &  (0.242) &  (0.224) \\
			&      & 0.7 & 500 &        -0.445 &    0.104 &   -0.216 &     -0.454 &    0.151 &   -0.174 &     -0.512 &    0.087 &   -0.208 \\
			&      &     &      &       (0.326) &  (0.407) &  (0.333) &    (0.270) &  (0.315) &  (0.270) &    (0.198) &  (0.264) &  (0.243) \\
			&      &     & 1000 &        -0.445 &    0.073 &   -0.147 &     -0.497 &    0.011 &   -0.178 &     -0.503 &    0.051 &   -0.167 \\
			&      &     &      &       (0.194) &  (0.256) &  (0.232) &    (0.197) &  (0.230) &  (0.211) &    (0.155) &  (0.203) &  (0.185) \\
			deep-1 & 0.5& 0.4 & 500 &        -0.799 &   -0.348 &   -0.064 &     -0.782 &   -0.412 &   -0.052 &     -0.751 &   -0.362 &     -0.1 \\
			&      &     &      &       (0.309) &  (0.447) &  (0.414) &    (0.252) &  (0.384) &  (0.391) &    (0.272) &  (0.366) &  (0.371) \\
			&      &     & 1000 &        -0.771 &   -0.414 &   -0.025 &     -0.812 &   -0.438 &   -0.037 &     -0.803 &   -0.446 &   -0.041 \\
			&      &     &      &       (0.210) &  (0.314) &  (0.304) &    (0.211) &  (0.277) &  (0.261) &    (0.192) &  (0.261) &  (0.241) \\
			&      & 0.7 & 500 &        -0.754 &   -0.398 &    -0.07 &     -0.775 &    -0.44 &   -0.128 &     -0.739 &   -0.361 &    0.028 \\
			&      &     &      &       (0.300) &  (0.380) &  (0.420) &    (0.280) &  (0.403) &  (0.357) &    (0.294) &  (0.344) &  (0.342) \\
			&      &     & 1000 &        -0.795 &   -0.437 &    0.018 &     -0.779 &    -0.39 &    0.006 &     -0.765 &   -0.423 &    0.004 \\
			&      &     &      &       (0.215) &  (0.286) &  (0.253) &    (0.192) &  (0.231) &  (0.253) &    (0.188) &  (0.230) &  (0.195) \\
			deep-2 & 0.5 & 0.4 & 500 &        -0.365 &   -0.287 &   -0.091 &     -0.374 &   -0.325 &    -0.13 &     -0.374 &     -0.3 &   -0.004 \\
			&      &     &      &       (0.320) &  (0.344) &  (0.393) &    (0.292) &  (0.313) &  (0.343) &    (0.268) &  (0.308) &  (0.333) \\
			&      &     & 1000 &        -0.408 &   -0.392 &   -0.061 &     -0.432 &   -0.388 &   -0.054 &     -0.422 &   -0.387 &   -0.024 \\
			&      &     &      &       (0.189) &  (0.200) &  (0.224) &    (0.180) &  (0.191) &  (0.203) &    (0.179) &  (0.190) &  (0.197) \\
			&      & 0.7 & 500 &        -0.353 &   -0.299 &   -0.042 &      -0.43 &   -0.372 &   -0.092 &     -0.409 &   -0.381 &   -0.086 \\
			&      &     &      &       (0.288) &  (0.286) &  (0.331) &    (0.264) &  (0.310) &  (0.288) &    (0.242) &  (0.231) &  (0.273) \\
			&      &     & 1000 &        -0.367 &   -0.326 &    -0.04 &      -0.41 &   -0.377 &    -0.06 &     -0.395 &   -0.364 &   -0.034 \\
			&      &     &      &       (0.195) &  (0.221) &  (0.211) &    (0.170) &  (0.183) &  (0.193) &    (0.186) &  (0.184) &  (0.184) \\\hline
		\end{tabular}
	}
\label{tbl:bias}
\end{table}

\begin{table}[]
				\caption{95\% coverage probability of $\bm{\beta}$ for CPH, PLAC and DPLC}
	\resizebox{\textwidth}{!}{
		\begin{tabular}{lllllllllllll}
			\hline &      &     & & \multicolumn{3}{c}{$\kappa=0.5$} & \multicolumn{3}{c}{$\kappa=1$} & \multicolumn{3}{c}{$\kappa=2$} \\\hline
			setting & $\rho$ & $p$ & $n$ &  CPH &   PLAC &   DPLC &        CPH &   PLAC &   DPLC &        CPH &   PLAC &   DPLC \\\hline
			linear & 0.5& 0.4 & 500 &        0.980 &  0.990 &  0.940 &      1.000 &  0.980 &  0.980 &      0.940 &  0.960 &  0.920 \\
			&      &     & 1000 &        0.970 &  0.970 &  0.910 &      0.929 &  0.960 &  0.949 &      0.940 &  0.950 &  0.970 \\
			&      & 0.7 & 500 &        0.920 &  0.920 &  0.920 &      0.970 &  0.960 &  0.960 &      0.980 &  0.980 &  0.980 \\
			&      &     & 1000 &        0.980 &  0.980 &  0.889 &      0.980 &  0.970 &  0.960 &      0.970 &  0.960 &  0.960 \\
			additive & 0.5 & 0.4 & 500 &        0.750 &  0.983 &  0.933 &      0.667 &  0.983 &  0.883 &      0.567 &  1.000 &  0.833 \\
			&      &     & 1000 &        0.317 &  0.950 &  0.850 &      0.333 &  0.950 &  0.817 &      0.217 &  0.867 &  0.867 \\
			&      & 0.7 & 500 &        0.600 &  0.983 &  0.867 &      0.617 &  0.983 &  0.900 &      0.500 &  0.950 &  0.900 \\
			&      &     & 1000 &        0.367 &  0.950 &  0.833 &      0.267 &  0.967 &  0.817 &      0.150 &  0.950 &  0.833 \\
			deep-1 & 0.5 & 0.4 & 500 &        0.280 &  0.880 &  0.980 &      0.200 &  0.870 &  0.950 &      0.230 &  0.900 &  0.950 \\
			&      &     & 1000 &        0.051 &  0.646 &  0.919 &      0.040 &  0.670 &  0.950 &      0.020 &  0.566 &  0.949 \\
			&      & 0.7 & 500 &        0.310 &  0.910 &  0.920 &      0.240 &  0.820 &  0.950 &      0.280 &  0.870 &  0.940 \\
			&      &     & 1000 &        0.051 &  0.636 &  0.929 &      0.030 &  0.690 &  0.950 &      0.000 &  0.500 &  0.970 \\
			deep-2 & 0.5 & 0.4 & 500 &        0.683 &  0.867 &  0.883 &      0.767 &  0.800 &  0.883 &      0.733 &  0.817 &  0.917 \\
			&      &     & 1000 &        0.533 &  0.617 &  0.900 &      0.383 &  0.517 &  0.917 &      0.383 &  0.567 &  0.950 \\
			&      & 0.7 & 500 &        0.817 &  0.933 &  0.950 &      0.650 &  0.783 &  0.933 &      0.617 &  0.767 &  0.950 \\
			&      &     & 1000 &        0.533 &  0.650 &  0.967 &      0.317 &  0.450 &  0.950 &      0.350 &  0.433 &  0.983 \\\hline
		\end{tabular}
	}
\label{tbl:cp}
\end{table}

\begin{table}[]
		\caption{Mean of the squared prediction errors evaluated on the test set for the CPH, PLAC and DPLC methods.}
	\resizebox{\textwidth}{!}{
		\begin{tabular}{ccccccccccccc}\hline
			&      &     &      & \multicolumn{3}{c}{$\kappa=0.5$} & \multicolumn{3}{c}{$\kappa=1$} & \multicolumn{3}{c}{$\kappa=2$} \\\hline
			setting & $\rho$ & $p$ & $n$ &           CPH &     PLAC &     DPLC &        CPH &     PLAC &     DPLC &        CPH &     PLAC &     DPLC \\\hline
			linear & 0.5 & 0.4 & 500 &         0.021 &    0.150 &    0.087 &      0.019 &    0.111 &    0.067 &      0.023 &    0.098 &    0.065 \\
			&      &     &      &       (0.018) &  (0.140) &  (0.084) &    (0.017) &  (0.072) &  (0.058) &    (0.023) &  (0.064) &  (0.055) \\
			&      &     & 1000 &         0.008 &    0.052 &    0.037 &      0.010 &    0.045 &    0.039 &      0.008 &    0.034 &    0.027 \\
			&      &     &      &       (0.006) &  (0.045) &  (0.032) &    (0.008) &  (0.028) &  (0.037) &    (0.006) &  (0.022) &  (0.019) \\
			&      & 0.7 & 500 &         0.022 &    0.127 &    0.124 &      0.017 &    0.095 &    0.071 &      0.015 &    0.062 &    0.047 \\
			&      &     &      &       (0.021) &  (0.109) &  (0.504) &    (0.014) &  (0.066) &  (0.056) &    (0.011) &  (0.035) &  (0.031) \\
			&      &     & 1000 &         0.008 &    0.044 &    0.041 &      0.008 &    0.033 &    0.033 &      0.007 &    0.022 &    0.022 \\
			&      &     &      &       (0.005) &  (0.024) &  (0.041) &    (0.006) &  (0.021) &  (0.027) &    (0.003) &  (0.008) &  (0.014) \\
			additive & 0.5 & 0.4 & 500 &         0.922 &    0.114 &    0.374 &      0.927 &    0.088 &    0.340 &      0.918 &    0.084 &    0.312 \\
			&      &     &      &       (0.028) &  (0.062) &  (0.091) &    (0.022) &  (0.040) &  (0.076) &    (0.022) &  (0.044) &  (0.062) \\
			&      &     & 1000 &         0.915 &    0.041 &    0.280 &      0.919 &    0.032 &    0.244 &      0.915 &    0.031 &    0.235 \\
			&      &     &      &       (0.023) &  (0.025) &  (0.042) &    (0.016) &  (0.013) &  (0.050) &    (0.019) &  (0.014) &  (0.043) \\
			&      & 0.7 & 500 &         0.921 &    0.095 &    0.368 &      0.925 &    0.072 &    0.311 &      0.922 &    0.055 &    0.287 \\
			&      &     &      &       (0.024) &  (0.057) &  (0.110) &    (0.016) &  (0.029) &  (0.049) &    (0.025) &  (0.024) &  (0.052) \\
			&      &     & 1000 &         0.914 &    0.038 &    0.251 &      0.925 &    0.029 &    0.222 &      0.914 &    0.024 &    0.214 \\
			&      &     &      &       (0.015) &  (0.021) &  (0.048) &    (0.015) &  (0.012) &  (0.047) &    (0.014) &  (0.009) &  (0.031) \\
			deep-1 & 0.5 & 0.4 & 500 &         0.984 &    0.335 &    0.124 &      0.983 &    0.312 &    0.090 &      0.985 &    0.310 &    0.076 \\
			&      &     &      &       (0.012) &  (0.081) &  (0.110) &    (0.013) &  (0.026) &  (0.052) &    (0.015) &  (0.022) &  (0.060) \\
			&      &     & 1000 &         0.979 &    0.290 &    0.060 &      0.981 &    0.298 &    0.056 &      0.979 &    0.301 &    0.033 \\
			&      &     &      &       (0.008) &  (0.016) &  (0.039) &    (0.009) &  (0.017) &  (0.042) &    (0.007) &  (0.018) &  (0.040) \\
			&      & 0.7 & 500 &         0.987 &    0.325 &    0.103 &      0.977 &    0.293 &    0.085 &      0.983 &    0.291 &    0.062 \\
			&      &     &      &       (0.014) &  (0.028) &  (0.063) &    (0.015) &  (0.020) &  (0.065) &    (0.010) &  (0.019) &  (0.119) \\
			&      &     & 1000 &         0.983 &    0.303 &    0.071 &      0.976 &    0.287 &    0.052 &      0.980 &    0.295 &    0.025 \\
			&      &     &      &       (0.008) &  (0.013) &  (0.123) &    (0.010) &  (0.017) &  (0.054) &    (0.007) &  (0.020) &  (0.017) \\
			deep-2 & 0.5 & 0.4 & 500 &         0.971 &    0.936 &    0.448 &      0.977 &    0.939 &    0.397 &      0.980 &    0.947 &    0.316 \\
			&      &     &      &       (0.023) &  (0.039) &  (0.243) &    (0.023) &  (0.036) &  (0.155) &    (0.014) &  (0.033) &  (0.132) \\
			&      &     & 1000 &         0.953 &    0.874 &    0.188 &      0.960 &    0.883 &    0.163 &      0.967 &    0.900 &    0.142 \\
			&      &     &      &       (0.012) &  (0.018) &  (0.059) &    (0.011) &  (0.019) &  (0.046) &    (0.010) &  (0.016) &  (0.043) \\
			&      & 0.7 & 500 &         0.974 &    0.932 &    0.361 &      0.989 &    0.951 &    0.323 &      0.981 &    0.924 &    0.248 \\
			&      &     &      &       (0.020) &  (0.031) &  (0.122) &    (0.023) &  (0.035) &  (0.130) &    (0.023) &  (0.034) &  (0.100) \\
			&      &     & 1000 &         0.960 &    0.886 &    0.173 &      0.972 &    0.898 &    0.154 &      0.966 &    0.877 &    0.120 \\
			&      &     &      &       (0.015) &  (0.022) &  (0.058) &    (0.011) &  (0.017) &  (0.054) &    (0.014) &  (0.020) &  (0.046) \\\hline
		\end{tabular}
	}
\label{tbl:mse}
\end{table}

\begin{table}[]
	\caption{$\mathrm{IMSE}$ evaluated on the test set for the CPH, PLAC and DPLC methods.}
	\resizebox{\textwidth}{!}{
		\begin{tabular}{ccccccccccccc}\hline
			&      &     &      & \multicolumn{3}{c}{$\kappa=0.5$} & \multicolumn{3}{c}{$\kappa=1$} & \multicolumn{3}{c}{$\kappa=2$} \\\hline
			setting & $\rho$ & $p$ & $n$ &           CPH &     PLAC &     DPLC &        CPH &     PLAC &     DPLC &        CPH &     PLAC &     DPLC \\\hline
			linear & 0.5 & 0.4 & 500 &         0.088 &    0.093 &    0.093 &      0.072 &    0.076 &    0.078 &      0.053 &    0.056 &    0.057 \\
			&      &     &      &       (0.002) &  (0.003) &  (0.004) &    (0.002) &  (0.003) &  (0.005) &    (0.001) &  (0.002) &  (0.003) \\
			&      &     & 1000 &         0.086 &    0.088 &    0.088 &      0.070 &    0.072 &    0.073 &      0.052 &    0.053 &    0.054 \\
			&      &     &      &       (0.001) &  (0.002) &  (0.003) &    (0.001) &  (0.002) &  (0.002) &    (0.001) &  (0.001) &  (0.002) \\
			&      & 0.7 & 500 &         0.093 &    0.098 &    0.098 &      0.083 &    0.087 &    0.088 &      0.062 &    0.065 &    0.066 \\
			&      &     &      &       (0.003) &  (0.004) &  (0.004) &    (0.001) &  (0.002) &  (0.004) &    (0.001) &  (0.001) &  (0.003) \\
			&      &     & 1000 &         0.091 &    0.093 &    0.094 &      0.081 &    0.083 &    0.084 &      0.062 &    0.063 &    0.063 \\
			&      &     &      &       (0.001) &  (0.001) &  (0.002) &    (0.001) &  (0.001) &  (0.002) &    (0.001) &  (0.001) &  (0.001) \\
			additive & 0.5 & 0.4 & 500 &         0.209 &    0.131 &    0.159 &      0.186 &    0.109 &    0.138 &      0.141 &    0.079 &    0.103 \\
			&      &     &      &       (0.004) &  (0.004) &  (0.008) &    (0.003) &  (0.003) &  (0.006) &    (0.003) &  (0.002) &  (0.006) \\
			&      &     & 1000 &         0.206 &    0.125 &    0.150 &      0.184 &    0.105 &    0.129 &      0.139 &    0.076 &    0.096 \\
			&      &     &      &       (0.003) &  (0.002) &  (0.004) &    (0.003) &  (0.001) &  (0.004) &    (0.002) &  (0.001) &  (0.003) \\
			&      & 0.7 & 500 &         0.220 &    0.134 &    0.166 &      0.202 &    0.116 &    0.146 &      0.163 &    0.088 &    0.115 \\
			&      &     &      &       (0.003) &  (0.003) &  (0.009) &    (0.003) &  (0.003) &  (0.005) &    (0.002) &  (0.002) &  (0.004) \\
			&      &     & 1000 &         0.217 &    0.128 &    0.154 &      0.199 &    0.112 &    0.137 &      0.161 &    0.086 &    0.107 \\
			&      &     &      &       (0.002) &  (0.001) &  (0.005) &    (0.002) &  (0.001) &  (0.005) &    (0.001) &  (0.001) &  (0.003) \\
			deep-1 & 0.5 & 0.4 & 500 &         0.231 &    0.129 &    0.082 &      0.214 &    0.117 &    0.070 &      0.184 &    0.100 &    0.057 \\
			&      &     &      &       (0.004) &  (0.004) &  (0.006) &    (0.003) &  (0.003) &  (0.006) &    (0.004) &  (0.003) &  (0.005) \\
			&      &     & 1000 &         0.228 &    0.123 &    0.076 &      0.212 &    0.114 &    0.066 &      0.181 &    0.096 &    0.052 \\
			&      &     &      &       (0.002) &  (0.003) &  (0.002) &    (0.003) &  (0.002) &  (0.004) &    (0.003) &  (0.002) &  (0.002) \\
			&      & 0.7 & 500 &         0.239 &    0.139 &    0.087 &      0.227 &    0.122 &    0.075 &      0.203 &    0.109 &    0.061 \\
			&      &     &      &       (0.004) &  (0.004) &  (0.006) &    (0.003) &  (0.003) &  (0.005) &    (0.003) &  (0.003) &  (0.003) \\
			&      &     & 1000 &         0.236 &    0.133 &    0.082 &      0.225 &    0.118 &    0.070 &      0.200 &    0.107 &    0.058 \\
			&      &     &      &       (0.002) &  (0.002) &  (0.003) &    (0.002) &  (0.002) &  (0.003) &    (0.002) &  (0.002) &  (0.002) \\
			deep-2 & 0.5 & 0.4 & 500 &         0.208 &    0.209 &    0.166 &      0.175 &    0.176 &    0.135 &      0.129 &    0.129 &    0.095 \\
			&      &     &      &       (0.004) &  (0.005) &  (0.014) &    (0.003) &  (0.004) &  (0.012) &    (0.003) &  (0.003) &  (0.007) \\
			&      &     & 1000 &         0.204 &    0.201 &    0.147 &      0.173 &    0.169 &    0.118 &      0.127 &    0.124 &    0.086 \\
			&      &     &      &       (0.002) &  (0.002) &  (0.005) &    (0.002) &  (0.003) &  (0.003) &    (0.002) &  (0.002) &  (0.003) \\
			&      & 0.7 & 500 &         0.218 &    0.217 &    0.169 &      0.195 &    0.196 &    0.148 &      0.150 &    0.148 &    0.106 \\
			&      &     &      &       (0.003) &  (0.004) &  (0.010) &    (0.003) &  (0.004) &  (0.011) &    (0.002) &  (0.003) &  (0.006) \\
			&      &     & 1000 &         0.216 &    0.211 &    0.154 &      0.192 &    0.190 &    0.135 &      0.148 &    0.143 &    0.098 \\
			&      &     &      &       (0.002) &  (0.002) &  (0.004) &    (0.001) &  (0.002) &  (0.004) &    (0.002) &  (0.002) &  (0.003) \\\hline
		\end{tabular}
	}
\label{tbl:imse1}
\end{table}

\addtolength{\textheight}{.5in}%

\section{Application}
\label{sec:app}

The Chinese Longitudinal Healthy Longevity Survey(CLHLS) is a follow-up survey of the elderly in China  organized by Peking University Healthy Aging and Development Research Center and the National Development Research Institute. This  longitudinal study covered 23 provinces across the country  with the elderly aged 65 and above and is the earliest and longest social science survey in China(1998-2018).

Its main objective is to identify the prevalent factors affecting the health and quanlity of life of the China's elderly. The questionnaire for the respondents includes the basic conditions of the elderly and their families, socio-economic background, self-assessment of health and quality of life, cognitive function, personality and psychological characteristics, daily activities, lifestyle, diseases and treatment, etc.

In this section, our primary purpose  is to assess the effectiveness of our  method in adjusting for non-linear covariate effects associated with cognitive impairment while still maintaining good interpretability of some covariates. We use the 2008 wave of the CLHLS, which  includes 5813 sujects. To determine the interval that brackets the time to cognitive impairment, the cognitive status is measured at each visit using the Chinese verision of MMSE, which includes 24 cognitive questions with scores ranging from 0-30. The right endpoint of this interval is considered to be the last visit time after which all MMSE scores are below 18, and the left endpoint is the previous visit time. The average length of the interval is 5.328 years if not right-censored. Following \cite{Lv2019}, we include continuous covariates age, years of education, boday mass index(BMI), systolic
blood pressures (SBP) and diastolic blood pressures (DBP) as $\bm{Z}$, and the binary covariates sex, exercise, diabetes and stroke or cardiovascular diease as $\bm{X}$. These covariates are summarized in Table 3 in Supplementary Material. We randomly divided the sample into a training set (80\%of the total) and a test set (the remaining 20\%). As in the simulation study, the DNN architecture is chosen from some pre-spcified candidates.

Similar to the simulation study, we compare the predictive power between the DPLC, CPH and PLAC in terms of the $\mathrm{IMSE}$.
  After fitting the model, the $\mathrm{IMSE}$ for DPLC is 0.098 while  for CPH and PLAC, it is 0.131 and 0.119, respectively. This shows that the performance of our approach is empirically superior to PLAC which is also superior to CPH. This can be intuitively explained by the fact that those covariate effects are highly nolinear and the more complex the sieve, the better the fit. The estimated coefficients of the linearly modelled covariates are shown in Table \ref{tbl:betapp}. From this table, it can be seen that being male rather than female and more exercise are significantly associated with a lower risk of cognitive impairment while diabetes and stroke or cardiovascular diease are not significant. These conclusions are consistent with those of \cite{Lv2019}.
%\begin{table}[]
%	\caption{$\mathrm{IMSE}_1$ and $\mathrm{IMSE}_2$ evaluated on the test set for the CPH, PLAC and DPLC methods for CLHLS.}
%	\begin{tabular}{llll}
%		\hline
%		&  CPH& PLAC & DPLC\\\hline
%%		$\mathrm{IMSE}_1$	& 0.121 & 0.115 &  0.095  \\
%		$\mathrm{IMSE}$	& 0.128 & 0.123 & 0.106 \\\hline
%	\end{tabular}
%\label{tbl:imseapp}
%\end{table}

\begin{table}[]
	\caption{estimation of $\bm{\beta}$ with exercise, diabetes and stoke or cardiovascular dieases}
	\begin{tabular}{@{}lccccc@{}}
		\hline 
		& EST & HR & SE & $Z$-value & $p$-value \\\hline
		Gender(Femal=1) & $0.148$    & $1.159$    & $0.019$     & $7.789$   & $0.000$               \\
		Exercise & $-0.283$   & $0.753$    & $0.023$     & $-12.304$ & $0.000$               \\
		Diabetes & $-0.025$   & $0.975$    & $0.020$     & $-1.25$  & $0.182$                 \\
		Stroke or cardiovascular dieases & $0.029$    & $1.029$    & $0.023$     & $1.26$   & $0.171$                 \\\hline
	\end{tabular}
	\label{tbl:betapp}
\end{table}

\section{Conclusion}
\label{sec:conc}
In this paper, we propose a partial linear Cox model with a deep neural network estimated nonparametric component for interval-censored failure time data. This model increases  predictive power compared to the partial linear additive Cox model(\cite{Lu2018}), while retaining interpretation for the covariates of  primary interest. The estimators are showed to converge at a rate independent of the nonparametric covariate dimensionality and the parametric estimator is rigorously proved to be asymptotically normal and semiparametric efficient. As shown in simulation studies, the proposed model significantly outperforms the linear Cox model and the partial linear additive Cox model with respect to both estimation and prediction when the nonparametric function is enormously complex.

%While DEEPSURV provides a fully nonparametric method for risk prediction and treatmenet recommendation, it suffers from the lack
%interpretability. In order to balance interpretability and prediction power, we propose a partial linear Cox model that allows part of all covariates to be nonparametric form.
This model is suitable for moderate sample sizes, but otherwise may encounter some problems due to the high non-convexity and complexity of DNN. With a small sample size, the optimization becomes unstable and one has to re-initialize for more often, while with a large sample, a large capacity DNN  is preferred and the optimization becomes quite time-consuming.

This work can be extended in several ways. For example, other types of semiparametric survival models such as the partial linear transformation model(\cite{Ma2005}) can be developed using the same procedure. Another interesting future work is to use deep convolutional neural networks(CNN, \cite{6795724}), a popular variant of DNN, to estimate the function $g$ when the nonparametric covariate is a high-dimensional image(\cite{Zhu2016}) or other types of unstructured data.

%\begin{align*}
%\mathcal{H}_r^{\alpha}(\mathcal{D}, M)=\bbrace{g:\mathcal{D}\mapsto \R: \sum_{\kappa:|\kappa|<\alpha}\norm{\partial^{\kappa} g}_{\infty}+\sum_{\kappa:|\kappa|=\alpha}\sup_{x,y\in \mathcal{D},x\ne y}\frac{|\partial^{\kappa}g(x)-\partial^{\kappa}g(y)|}{\norm{x-y}_{\infty}^{\alpha-\alpha}}\leq M}
%\end{align*}

\bibliographystyle{chicago}

\bibliography{bib}
\end{document}